\documentclass{article}
\usepackage{graphicx} 
\usepackage{xspace}
\newcommand{\CG}[0]{ChatGPT\xspace}
\newcommand{\tool}[0]{$\mathtt{GPT}$$\bullet$$\mathtt{LODS}$\xspace}

\usepackage{graphicx}
\usepackage{hyperref}

\begin{document}
\title{Using Multiple RDF Knowledge Graphs for
Enriching ChatGPT Responses}

\author{
Michalis Mountantonakis and Yannis Tzitzikas\\\\
Institute of Computer Science - FORTH-ICS, Greece\\
	and
	Computer Science Department - University of Crete, Greece \\
\{mountant, tzitzik\}@ics.forth.gr
}
\date{April, 2023}
\maketitle
\begin{abstract}
There is a recent trend for using the novel Artificial Intelligence \CG chatbox, which provides detailed responses and articulate answers across many domains of knowledge. However, in many cases it returns plausible-sounding but incorrect or inaccurate responses, whereas it does not provide evidence. Therefore, any user has to further search for checking the accuracy of the answer or/and for finding more information about the entities of the response. At the same time there is a high proliferation of  RDF Knowledge Graphs (KGs) over any real domain, that offer high quality structured data. For enabling the combination of \CG and RDF KGs, we present a research prototype, called \tool, which is able to enrich any \CG response with more information from hundreds of RDF KGs. In particular, it identifies and annotates each entity of the response with statistics and hyperlinks to LODsyndesis KG (which contains integrated data from 400 RDF KGs and over 412 million entities). In this way, it is feasible to enrich the content of entities and to perform fact checking and validation for the facts of the response at real time.\newline \newline

\textbf{Demo Video}: \url{https://youtu.be/H30bSv9NfUw}.

\textbf{Demo URL}: 
\url{https://demos.isl.ics.forth.gr/GPToLODS}.
\end{abstract}

\section{Introduction}
\CG\ is a novel Artificial Intelligence (AI) chatbox (\url{https://chat.openai.com/}), which is built on GPT-3.5 and GPT-4 families of large language models (LLMs) \cite{brown2020language}, and provides detailed responses and human-like answers across many domains of knowledge. However, it has not been designed to store or retrieve facts, e.g., like a relational database or a Knowledge Graph (KG). For this reason, in many cases it
returns plausible-sounding but incorrect or inaccurate responses \cite{van2023chatgpt}. Thereby, it is hard for the user to check
the validity of the answers returned by \CG. 
Sometimes the responses contain entities
that do not exist,
URLs that are wrong, facts that cannot be verified, outdated data,  and many others.   On the contrary, there are available numerous RDF KGs \cite{hogan2021knowledge} (e.g., DBpedia \cite{lehmann2015dbpedia}, Wikidata \cite{vrandevcic2014wikidata}, YAGO \cite{rebele2016yago}, etc), that provides high quality structured data (that are updated at least periodically), by using Linked Data techniques \cite{bizer2011linked}.

\begin{figure}[t]
    \centering
\fbox{\includegraphics[width=0.9\linewidth]{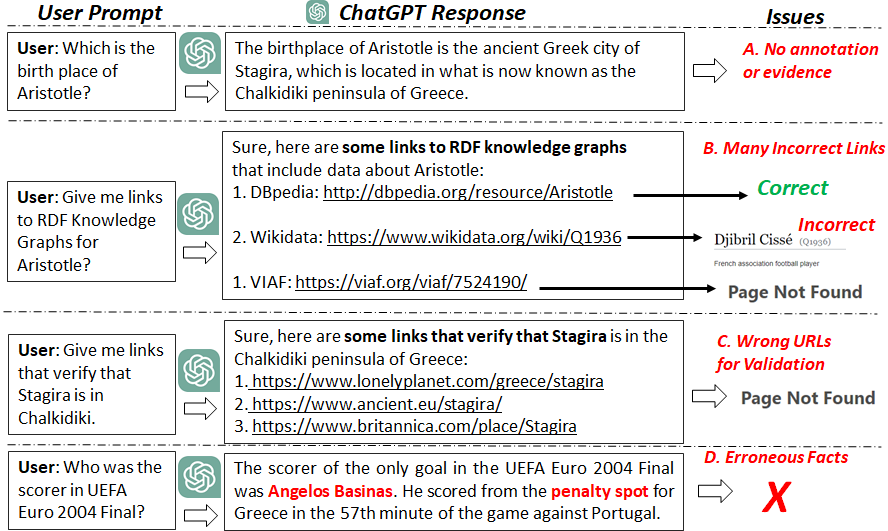}}
\caption{ChatGPT issues through real conversations (March 27, 2023)}
    \label{fig:issues}
\vspace{-2mm}
\end{figure}

Concerning some key issues of \CG (that can be improved by using KGs), Fig. \ref{fig:issues} shows a real conversation with \CG\ (in March 27, 2023). First,  we asked about the birth place of Aristotle (issue A), and the output was a plain text, without evidence or annotations. Afterwards, we desired to find available RDF links for Aristotle (issue B), and only one of them was correct, whereas the URI of Wikidata refers to a completely different entity. Then, we asked about sources verifying that Stagira is located in Chalkidiki, and all the returned URIs were invalid (issue C).  In the last case, we asked a  question from a different domain, i.e., ``Who scored the goal in UEFA Euro 2004 Final?", and we retrieved erroneous facts  (issue D);  its response was ``Angelos Basinas from the penalty", however the correct answer is ``Angelos Charisteas with a header".

The objective of this paper is to aid the above issues, however, there are several challenges that should be tackled. Indeed, it is quite difficult to check the validity of entities and URIs, since
it requires access to numerous KGs, sources
and resources in general.  For tacking these challenges,  we demonstrate the research prototype \tool,
that enables the user to make a question,
and instead of getting the raw answer 
from \CG,
it retrieves 
the \CG response
annotated with the identified entities,
and with relevant data 
about these entities
(URIs, facts and KGs),
For making this feasible at real time, we
exploit LODsyndesis suite of services \cite{mountantonakis2021services,mountantonakis2020content} including an  Information Extraction (IE) service \cite{mountantonakis2022linking}. These  services rely on LODsyndesis KG; a large KG, equipped with special indexes and algorithms, that has integrated 2 billion facts for 412 million entities, from 400 real RDF KGs. In this way, the user is able 
to speed up the validation of the \CG response
and to retrieve more information. Concerning the novelty, to the best of our knowledge it is the first system offering annotation and linking of the \CG\ responses to hundreds of KGs.

The rest of this demo paper is organized as follows; \ref{sec:rel} describes the related work, \ref{sec:process} presents the process and the use cases and \ref{sec:conc} concludes the paper.

\noindent



\section{Related Work}
\label{sec:rel}
First, concerning  \CG\ and KGs, \cite{omar2023chatgpt} provides a comparison for the Question Answering task and they concluded that \CG\ can have high precision in general knowledge domains but very low scores in unseen domains, compared to a KG-based approach.  Regarding IE tools for Entity Recognition, there are available approaches from several areas, i.e., from Natural Language Processing (NLP)\cite{manning2014stanford}, from KGs \cite{mendes2011dbpedia,piccinno2014tagme,moro2014multilingual}, and from Neural Networks \cite{van2020rel}. However, these tools link the entities to a single KG, and for this reason \tool\ uses the machinery of LODsyndesisIE \cite{mountantonakis2022linking}, which combines tools from  NLP and KG, and links the entities to 400 RDF KGs.
Regarding the novelty, to the best of our knowledge there is no other related system that annotates the response of \CG\ and provides links and services using hundreds of KGs at real time.

\begin{figure}[t]
    \centering
\includegraphics[width=0.8\linewidth]{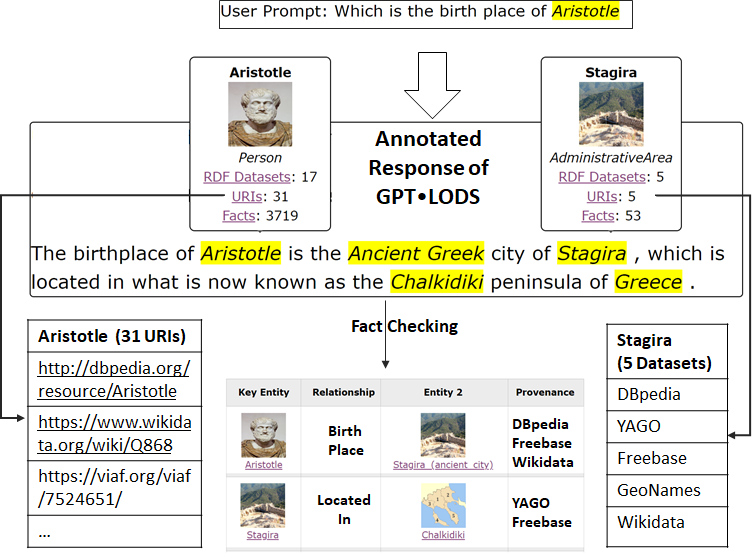}
\caption{Screenshots from \tool\ research prototype}
    \label{fig:lodgpt}
\vspace{-5mm}
\end{figure}

\section{The Process of \tool and Use Cases}
\label{sec:process}


First, the name of the prototype \tool\ comes from the mathematical notation for function composition, i.e.
($GPT$ $\bullet$ $LODS$)($x$) =$LODS(GPT(x))$, where $LODS$ comes from LODSyndesis \cite{mountantonakis2020content}. Concerning the process (see Fig. \ref{fig:lodgpt}), the user submits through the web application a question (in english),  \tool\ sends the question to the   \CG API and retrieves the response.  Afterwards,  it applies the machinery of LODsyndesisIE
(described in \cite{mountantonakis2022linking}) for recognizing the entities of the response, i.e., it 
combines widely used Entity Recognition tools
(i.e., DBpedia Spotlight \cite{mendes2011dbpedia}, WAT \cite{piccinno2014tagme} and Stanford CoreNLP\cite{manning2014stanford}) for recognizing the entities. The next step is to further process the response for creating the annotations and for adding statistics and links to LODsyndesis KG for each recognized entity, and finally it returns the annotated response to the user. Fig. \ref{fig:lodgpt} shows that by clicking on each entity,  one can see its name, image, type and statistics, such as 
the number of its RDF KGs, URIs and facts that occur in LODSyndesis. Moreover, by clicking on the links the user can browse (or download) all this data (e.g., the list of all the URIs of each entity). Finally, several other services are offered, including a  fact checking service (see Fig. \ref{fig:lodgpt}) that shows all the relations between any pair of entities of the response (for fact validation). 
\newline\indent \textbf{Scenario \& Use Cases.}
Below, we see a scenario which is planned to be also demonstrated in the conference. The target audience can be any researcher of AI area, since \tool\ (\url{https://demos.isl.ics.forth.gr/GPToLODS}) is a research prototype that combines tools and techniques of several AI components, including LLMs, NLP,  and Knowledge Representation and Reasoning (Linked Data and KGs).  The scenario is about the user questions (and needs) that  are shown in Fig. \ref{fig:issues}, i.e.,  starting with the question ``Which is the birth Place of Aristotle". For tackling the needs of this scenario,
the results include:
the  annotated entities,
 related information
(identifiers, facts and datasets) about these entities and validation of facts.
The scenario can be accessed in an online video (\url{https://youtu.be/H30bSv9NfUw}), which presents the issues of Fig. \ref{fig:issues} and how they can be solved through the following use cases. 
\newline\indent \textbf{Use Case 1. Annotation, Evidence and Linking.} 
This refers to the issues A and B of Figure \ref{fig:issues}, i.e., by having the annotation of the entities of the response, we are able to find more information (links, datasets and facts) for each of the entities of the response. Moreover, we can have access to the correct RDF links for each entity, e.g., see in the left part of Fig. \ref{fig:lodgpt}, that we retrieved the correct Wikidata and VIAF link for Aristotle (e.g., \url{https://www.wikidata.org/wiki/Q868}), and in total 31 URIs.
\newline\indent \textbf{Use Case 2. Fact Validation and Correct Answer.} 
In Fig. \ref{fig:issues} we can see that a \CG\ response can either provide wrong links for validation or even erroneous facts (issues C,D). In the first case, \tool\ can be used for fact validation, e.g., in Fig. \ref{fig:lodgpt} it verified 2 facts of the response from popular KGs, like DBpedia and Wikidata. Regarding issue D, \tool\ will not find the erroneous fact in LODsyndesis KG. For finding the correct answer, the user can further browse all the facts of an entity (by clicking on the  link).
\newline\indent \textbf{Use Case 3. Dataset Discovery and Enrichment.}
The user can discover all the datasets of each entity (e.g., see the lower right part of Fig. \ref{fig:lodgpt}), and all (or a part of) the facts of that entity in the KGs that are included in LODsyndesis. This can be useful for enriching the available content of each entity, e.g., for creating an application, a data warehouse \cite{mountantonakis2019large}, for performing an analysis, etc.
\vspace{-1mm}
\section{Concluding Remarks}
\label{sec:conc}
In this paper, we presented the research prototype \tool, which enables the real time annotation and linking of a \CG\ response to hundreds of RDF KGs, the enrichment of its entities and the validation of its facts. 
As a future work, we plan to improve the GUI, and the fact checking service by performing relation extraction, to offer a REST API and to support multilinguality.

\bibliographystyle{splncs04}
\bibliography{samplepaper}

\begin{thebibliography}{10}
\providecommand{\url}[1]{\texttt{#1}}
\providecommand{\urlprefix}{URL }
\providecommand{\doi}[1]{https://doi.org/#1}

\bibitem{bizer2011linked}
Bizer, C., Heath, T., Berners-Lee, T.: Linked data: The story so far. In:
  Semantic services, interoperability and web applications: emerging concepts,
  pp. 205--227. IGI global (2011)

\bibitem{brown2020language}
Brown, T., Mann, B., Ryder, N., Subbiah, M., Kaplan, J.D., Dhariwal, P.,
  Neelakantan, A., Shyam, P., Sastry, G., Askell, A., et~al.: Language models
  are few-shot learners. Advances in neural information processing systems
  \textbf{33},  1877--1901 (2020)

\bibitem{van2023chatgpt}
van Dis, E.A., Bollen, J., Zuidema, W., van Rooij, R., Bockting, C.L.: Chatgpt:
  five priorities for research. Nature  \textbf{614}(7947),  224--226 (2023)

\bibitem{hogan2021knowledge}
Hogan, A., Blomqvist, E., Cochez, M., d’Amato, C., Melo, G.d., Gutierrez, C.,
  Kirrane, S., Gayo, J.E.L., Navigli, R., Neumaier, S., et~al.: Knowledge
  graphs. ACM Computing Surveys (CSUR)  \textbf{54}(4),  1--37 (2021)

\bibitem{lehmann2015dbpedia}
Lehmann, J., Isele, R., Jakob, M., Jentzsch, A., Kontokostas, D., et~al.:
  Dbpedia--a large-scale, multilingual knowledge base extracted from wikipedia.
  Semantic web  \textbf{6}(2),  167--195 (2015)

\bibitem{manning2014stanford}
Manning, C.D., Surdeanu, M., Bauer, J., Finkel, J.R., Bethard, S., McClosky,
  D.: The stanford corenlp natural language processing toolkit. In: Proceedings
  of 52nd annual meeting of the association for computational linguistics:
  system demonstrations. pp. 55--60 (2014)

\bibitem{mendes2011dbpedia}
Mendes, P.N., Jakob, M., Garc{\'\i}a-Silva, A., Bizer, C.: Dbpedia spotlight:
  shedding light on the web of documents. In: Proceedings of the 7th
  international conference on semantic systems. pp.~1--8 (2011)

\bibitem{moro2014multilingual}
Moro, A., Cecconi, F., Navigli, R.: Multilingual word sense disambiguation and
  entity linking for everybody. In: ISWC (Posters \& Demos). pp. 25--28.
  Citeseer (2014)

\bibitem{mountantonakis2021services}
Mountantonakis, M.: Services for Connecting and Integrating Big Numbers of
  Linked Datasets, vol.~50. IOS Press (2021)

\bibitem{mountantonakis2019large}
Mountantonakis, M., Tzitzikas, Y.: Large-scale semantic integration of linked
  data: A survey. ACM Computing Surveys (CSUR)  \textbf{52}(5),  1--40 (2019)

\bibitem{mountantonakis2020content}
Mountantonakis, M., Tzitzikas, Y.: Content-based union and complement metrics
  for dataset search over rdf knowledge graphs. Journal of Data and Information
  Quality (JDIQ)  \textbf{12}(2),  1--31 (2020)

\bibitem{mountantonakis2022linking}
Mountantonakis, M., Tzitzikas, Y.: Linking entities from text to hundreds of
  rdf datasets for enabling large scale entity enrichment. Knowledge
  \textbf{2}(1),  1--25 (2022)

\bibitem{omar2023chatgpt}
Omar, R., Mangukiya, O., Kalnis, P., Mansour, E.: Chatgpt versus traditional
  question answering for knowledge graphs: Current status and future directions
  towards knowledge graph chatbots. arXiv preprint arXiv:2302.06466  (2023)

\bibitem{piccinno2014tagme}
Piccinno, F., Ferragina, P.: From tagme to wat: a new entity annotator. In:
  Proceedings of the first international workshop on Entity recognition \&
  disambiguation. pp. 55--62 (2014)

\bibitem{rebele2016yago}
Rebele, T., Suchanek, F., Hoffart, J., Biega, J., Kuzey, E., Weikum, G.: Yago:
  A multilingual knowledge base from wikipedia, wordnet, and geonames. In: The
  Semantic Web--ISWC 2016: 15th International Semantic Web Conference, Kobe,
  Japan, October 17--21, 2016, Proceedings, Part II 15. pp. 177--185. Springer
  (2016)

\bibitem{van2020rel}
Van~Hulst, J.M., Hasibi, F., Dercksen, K., Balog, K., de~Vries, A.P.: Rel: An
  entity linker standing on the shoulders of giants. In: Proceedings of the
  43rd International ACM SIGIR Conference on Research and Development in
  Information Retrieval. pp. 2197--2200 (2020)

\bibitem{vrandevcic2014wikidata}
Vrande{\v{c}}i{\'c}, D., Kr{\"o}tzsch, M.: Wikidata: a free collaborative
  knowledgebase. Communications of the ACM  \textbf{57}(10),  78--85 (2014)

\end{thebibliography}

\end{document}